\documentclass[prc,twocolumn,showpacs,nofootinbibs,superscriptaddress]{revtex4}
\usepackage{epsf}
\usepackage{amsmath}
\usepackage{xcolor}

\def\beq{\begin{equation}}
\def\eeq{\end{equation}}
\def\bea{\begin{eqnarray}}
\def\eea{\end{eqnarray}}
\def\beqa{\begin{equation}\begin{array}{l}}
\def\eeqa{\end{array}\end{equation}}
\def\eqlab#1{\label{eq:#1}}
\def\figlab#1{\label{fig:#1}}
\def\tablab#1{\label{tab:#1}}

\def\Eqref#1{Eq.~(\ref{eq:#1})}

\def\fref#1{\ref{fig:#1}}
\def\Figref#1{Fig.~\ref{fig:#1}}
\def\tabref#1{\ref{tab:#1}}

\def\sla#1{#1 \!\!\!\! \slash\,}

\def\slad{\partial \hspace{-2.1mm} \slash}

\def\slaa{a \hspace{-2mm} \slash}
\def\half{\mbox{\small{$\frac{1}{2}$}}}

\def\barr{\left(\begin{array}{c}}
\def\earr{\end{array}\right)}
\def\bmat{\left(\begin{array}{cc}}
\def\emat{\end{array}\right)}
\def\al{\alpha}
\def\be{\beta}
\def\ga{\gamma} 
 \def\De{\Delta}\def\vDe{\varDelta}
\def\veps{\varepsilon}  \def\eps{\epsilon}

\def\la{\lambda}

\def\si{\sigma} 
  
\def\w{\omega}

\def\pa{\partial}

\def\pa{\partial}

\def\nn{\nonumber}

\def\cO{\mathcal{O}}

\def\lag{{\mathcal L}}

\def\mathscr{\mathcal}

\def\3d{3-D}

\def\ol#1{\overline{#1}}

\begin{document}

\title{Manifestly-covariant chiral PT calculation of nucleon Compton scattering}
\author{Vadim Lensky}
\email{lensky@ect.it}

\affiliation{European Centre for Theoretical Studies in Nuclear Physics and Related Areas (ECT*), \\
Villa Tambosi, Villazzano (Trento),
I-38050 TN, Italy}

\affiliation{Institute for Theoretical and Experimental Physics, 117218 Moscow, Russia\footnote{On leave of absence.}}

\author{Vladimir Pascalutsa}
%\email{vlad@ect.it}

\affiliation{European Centre for Theoretical Studies in Nuclear Physics and Related Areas (ECT*), \\
Villa Tambosi, Villazzano (Trento),
I-38050 TN, Italy}

\affiliation{Institut f\"ur Kernphysik, Johannes Gutenberg Universit\"at, Mainz D-55099, Germany}

\date{\today}

\begin{abstract}
We compute the Compton scattering off the nucleons in the
framework of manifestly covariant baryon chiral perturbation
theory (B$\chi$PT). The results for observables
differ substantially from the corresponding calculations in heavy-baryon
chiral perturbation theory (HB$\chi$PT), most appreciably in
the forward kinematics. We verify that the covariant $p^3$ result
fulfills the forward-Compton-scattering sum rules. We also explore the effect of 
the $\Delta$(1232) resonance at order $p^4/\vDe$, with $\vDe\approx 300$ MeV,
the resonance excitation energy.
We find that the substantial effect of the $\Delta$-excitation on the nucleon polarizabilities
can naturally be accommodated in the 
manifestly covariant calculation.
\end{abstract}


\maketitle
\thispagestyle{empty}

{\it 1. Introduction}

Compton scattering off nucleons
is a unique tool to study the structure
and the e.m.\ properties of the nucleon.  
The e.m.\ properties are probed
already at very low energies (soft photons), 
where the process depends
on the static e.m.\ moments of the target, the 
charge and magnetic moment in the case of the nucleon~\cite{Low:1954kd}.
To probe the structure, the energy of the incident
photons must be sufficiently high, their wavelength to be
comparable with the nucleon size. 

From a pedestrian point of view the nucleon consists of a quark core
surrounded by a cloud of pions. The quark core is confined
to a radius of less than a fermi, while the pion cloud can extend
to as much as 1.5 fm. And indeed, at energies of around
100 MeV and above, the effects of the nucleon structure
 become significant enough to be detected in
modern Compton-scattering 
experiments~\cite{Federspiel:1991yd,Zieger:1992jq,Hal93,MacG95,MAMI01} 
(see Refs.~\cite{Drechsel:2002ar,Schumacher:2005an} for review). 
More specifically, it has been possible to detect the
nucleon {\em polarizabilities}, with 
the result shown in the PDG column of Table~\tabref{albe}.
These values clearly point out the composite structure of the nucleon
(for a classical pointlike object polarizabilities vanish). 
More insights come from studying these quantities in chiral
perturbation theory ($\chi$PT), an effective theory
of the low-energy strong interaction~\cite{Weinberg:1978kz,Gasser:1983yg}.

In $\chi$PT, the leading-order result for the nucleon
polarizabilities is a prediction, in the sense that the
low-energy constants (LECs) associated with polarizabilities
do not enter until the next order. The whole result is given by
a few chiral loops, quantified in terms of well-known parameters,
such the nucleon and pion masses ($M_N$ and $m_\pi$) and the
pion-nucleon coupling constant, $g_{\pi NN}$.
The first calculation of polarizabilities in $\chi$PT, carried out
by Bernard, Kaiser and Mei{\ss}ner~\cite{Bernard:1991rq},
at leading order yields the values shown in the
B$\chi$PT ${\cal O}(p^3)$ column of Table~\tabref{albe}.

This first calculation has been carried out in what now is called
{\it manifestly Lorentz-covariant} baryon chiral perturbation
theory (B$\chi$PT), to distinguish it from the heavy-baryon
chiral perturbation theory (HB$\chi$PT). 
The heavy-baryon formalism was carried over from
the heavy-quark QCD to $\chi$PT~\cite{JeM91a} in order to cure
the problems with chiral power counting, which B$\chi$PT had apparently
had~\cite{GSS89}. Incidentally, the HB$\chi$PT result (which is easily
obtained from~\cite{Bernard:1991rq} by 
keeping only the leading in $m_\pi/M_N$ term)
happens to agree with experiment much better, see Table~\tabref{albe}. 
Subsequently, more sophisticated
analyses of Compton scattering in HB$\chi$PT 
followed~\cite{BKM,McGovern:2001dd,Beane:2004ra}, 
while the original result~\cite{Bernard:1991rq} received less attention.

\begin{table}[t]
\begin{tabular}{||c|c|c||c|c||c||}
\hline\hline
&  \multicolumn{2}{|c||}{HB$\chi$PT} & 
\multicolumn{2}{|c||}{B$\chi$PT} & PDG \\
\cline{2-5} 
 &  ${\cal O}(p^3)$ \cite{BKM}&  ${\cal O}(\eps^3) $ \cite{Hemmert:1996rw}& $ {\cal O}(p^3)$ \cite{Bernard:1991rq}\footnote{The values differ from the ones quoted in
Ref.~\cite{Bernard:1991rq} because of the value of the $\pi NN$ coupling
constant: $g_{\pi NN}=13.4$ in \cite{Bernard:1991rq} versus $g_{\pi NN}=g_A M_N/f_\pi \simeq 12.9$ in this work and in~\cite{BKM}.}  &$ {\cal O}(p^4/\vDe) $&  \cite{PDG2006} \\
\hline
$\alpha^{(p)}$  & 12.2 & 20.8 & 6.8 & 10.8 & $12.0 \pm 0.6$\\
$\beta^{(p)}$ & 1.22 & 14.7 & $-1.8$ & 2.9 & $1.9\pm 0.5 $ \\ 
\hline\hline
\end{tabular}
\caption{The electric ($\al$) and magnetic ($\be$) polarizabilities
of the proton in units of $10^{-4}\,$fm$^3$. The last column quotes
the Particle Data Group compilation of experimental results, while
the first two represent the predictions of
the heavy-baryon  and manifestly-covariant $\chi$PT, respectively.
}
\tablab{albe}
\end{table}

More recently, however, Gegelia {\it et al.}~\cite{Gegelia:1999gf,Fuchs:2003qc}
realized that B$\chi$PT does not have a problem with power counting {\it per~se}. 
The pieces that
were thought to violate the power counting are compensated 
(renormalized)
by the available LECs and therefore do not alter the physics. 
At the same time, Becher and Leutwyler~\cite{BL99} point out that the difference
between B$\chi$PT and HB$\chi$PT result can be unnaturally large due
to the presence of physical cuts and other non-analytic structures
which do not easily admit a polynomial expansion.
Also, 
it has been noted that HB$\chi$PT
is incompatible with the dispersion relations 
and sum rules~\cite{Lvov:1993ex,Pascalutsa:2004ga,Holstein:2005db}.
Finally, the effect of the $\Delta$ excitation in Compton scattering
cannot be accommodated
in the HB framework in any natural way~\cite{Hildebrandt:2003fm, Pascalutsa:2003zk}
(see also the HB$\chi$PT column of Table~\tabref{albe}).
All these observations
make a strong case for adopting the manifestly-covariant formalism in favor
of the heavy-baryon one. The present work is the first step towards this goal.

In this letter we present the calculation of Compton scattering
in B$\chi$PT, to orders $p^3$ and $p^4/\De$. In Sect.~2 we consider the graphs needed 
to be computed and explain how we compute them. 
In Sect.~3, we verify that
 the B$\chi$PT result is consistent with the Compton-scattering sum rules.
We also elaborate on the impact that chiral symmetry makes 
on nucleon polarizabilities.
In Sect.~4 we show the results for the Compton scattering cross-sections.
We conclude with Sect.~5. 

\medskip

 {\it 2. Chiral loops and Lagrangians}
 
The chiral power counting for the Compton amplitude in $\chi$PT
with pions and nucleons is reviewed in~\cite{Bernard:1991rq,BKM}. 
Reference~\cite{Bernard:1991rq} also shows the one-loop graphs that arise at $\cO(p^3)$,
even if it only considers the contribution of those graphs to the scalar polarizabilities.

The chiral expansion for the Compton amplitude begins at order $p^2$, where it
receives contributions from the Born graphs only. At $\cO(p^3)$ there are
contributions from a number of loop graphs displayed in \Figref{loops} and the 
Wess-Zumino-Witten (WZW) anomaly graph.
The expression for the Born and WZW-anomaly contributions
can be found in, e.g.,~\cite{Pascalutsa:2003zk}, while here we focus on the loop graphs. 

Note that the graphs in \Figref{loops} are not exactly the same as found in~\cite{Bernard:1991rq},
however they will lead to exactly the same result. This is because we use a different,
albeit {\it equivalent}, in the sense of `equivalence theorem', form of the Lagrangian.
To explain that, let us consider the leading-order chiral Lagrangian
for the nucleon:
\beq
\lag^{(1)}_N = \ol N\,( i \sla{D} -{M}_{N} +  g_A \,  
\slaa\,\ga_5 )\, N\,,
\eqlab{Nlagran}
\eeq
where $N$ denotes the isodoublet Dirac field of the nucleon,
$M_N$ is the nucleon mass and $g_A$ is the axial-coupling 
constant, both taken at their chiral-limit value.
Furthermore, the chiral covariant derivative is given by
\beq 
D_\mu N = \pa_\mu N  + i v_\mu N \,,
\eeq  
whereas the vector and axial-vector fields above are defined 
in terms of the pion field, $\pi^a(x)$, as
\begin{subequations}
\bea
v_\mu & \equiv & \half\, \tau^a v_\mu^a(x) = \frac{1}{2i} \left(u \,
\pa_\mu u^\dagger+u^\dagger \pa_\mu  u \right), \\
a_\mu & \equiv & \half \,\tau^a a^{\,a}_\mu(x) =
  \frac{1}{2i} \left(u^\dagger \,
\pa_\mu  u- u \,\pa_\mu  u^\dagger \right) , 
\eea
\eqlab{currents}
\end{subequations}
with $u=\exp(i\pi^a \tau^a/2f )$, an $SU(2)$ matrix, and $f$ the pion decay constant in 
the chiral limit.
 
Let us now consider a redefinition of the nucleon
field, $N\to \xi N$, where the field $\xi$ has the following form:
\beq
\xi = \exp\left(\frac{ig_A\,\pi^a \tau^a}{2f}\ga_5 \right)\,.
\eeq
Our first-order chiral Lagrangian then becomes:\footnote{In our conventions
$\ga_5^\dagger= \gamma_5$, hence 
$\xi^\dagger=\exp(-ig\pi^a \tau^a\ga_5/2f_\pi)$, $\xi \xi^\dagger =1$.
Note also that $\ol N\to \ol N \xi$, and $\xi\ga^\mu\xi =\ga^\mu$.}
\bea
\eqlab{Nlagran2}
{\lag_N'}^{(1)}& = & \ol N\,\xi\,( i \sla{D} -{M}_{N} +  g_A \,  
\slaa\,\ga_5 )\, \xi\,N\,\nn\\
&=& \ol N\,( i \slad -{M}_{N} )\, N + M_N \,\ol N \,(1-\xi^2) N \\
 && + \ol N \, (  \xi \,i \slad\,\xi - \xi\,{v\!\!\! \slash} \,\xi 
+  g_A \, \xi\, \slaa\,\ga_5\,\xi ) \, N \,.\nn
\eea

\begin{figure}[tb]
\centerline{\epsfclipon   \epsfxsize=8.5cm%
  \epsffile{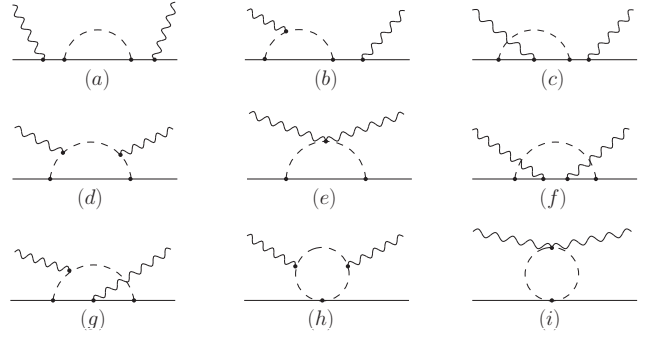} 
}
\caption{
The loop graphs evaluated in this work. Graphs obtained from these by
crossing and time-reversal are not shown, but are evaluated too. 
}
\figlab{loops}
\end{figure}

Both Lagrangians, \Eqref{Nlagran} and \Eqref{Nlagran2},
are equivalent, however, may have drastically different forms when
expanded in the pion field. For the one-loop contributions to Compton scattering
it is sufficient to expand up to the second
order in the pion field: 
\begin{subequations}
\bea
v_\mu &=&  \frac{1}{4f^2}\, \tau^a\veps^{abc} 
\pi^b\,\pa_\mu\pi^c+ O(\pi^3),\\
a_\mu &=& \frac{1}{2f}\tau^a\pa_\mu\pi^a+ O(\pi^3),\,\\
\xi &=& 1+ \frac{i g_A}{2f} \tau^a \pi^a \ga_5 - \frac{g_A^2}{8f^2}\pi^2 
+ O(\pi^3).
\eea
\end{subequations}
The original and the redefined Lagrangians take, respectively,
the following form:
\bea
\lag^{(1)}_N & = & \ol N\,\left( i \slad -{M}_{N}
+ \frac{g_A}{2f} \tau^a \slad\,\pi^a\ga_5 \right. \nn\\
&& - \left. \frac{1}{4f^2} \, \tau^a \veps^{abc}  \pi^b\,\slad \,\pi^c
\right)\, N  + O(\pi^3) \,,
\eqlab{expNlagran}\\
{\lag'}_N^{(1)} & = &  \ol N\,\left( i \slad -{M}_{N}\, 
-\, 
i\, \frac{ g_A}{f} M_N \tau^a\pi^a\ga_5 +  \frac{g_A^2}{2f^2} M_N \pi^2
\right. \nn\\
&& \left. -\frac{(g_A-1)^2}{4f^2} \, \tau^a\veps^{abc}  \pi^b\,\slad \,\pi^c
\right)\, N\,
 + O(\pi^3)\,.
\eqlab{expNlagran2}
\eea
The major difference between the two forms 
is that the pseudovector $\pi NN$ coupling is transformed into a pseudoscalar one, while
the Weinberg-Tomozawa $\pi\pi NN$ term is
replaced by an isoscalar term akin to the remains of an integrated-out $\sigma$-meson
in the linear sigma model. The new isovector $\pi\pi NN$ term, proportional to $(g_A-1)^2$ does not
give any contribution to Compton amplitude at one-loop level.

Also, at the order that loops are considered, the photon couples only minimally, i.e., to the electric charge of the
pion and nucleon.\footnote{Not in the Born graphs, where 
the anomalous magnetic moment of the nucleon appears too at this order.} 
Now that the pion couples to the nucleon via pseudoscalar coupling, there is no Kroll-Ruderman ($\gamma \pi NN$)
term arising, and hence the number of one-loop graphs is severely reduced.
 
Thus, the loop graphs shown in \cite{Bernard:1991rq} with couplings from the Lagrangian \Eqref{expNlagran}
transform to the graphs shown in \Figref{loops} with the couplings from \Eqref{expNlagran2}.
We have checked explicitly that the two sets
of one-loop diagrams give identical expressions for the Compton amplitude.

The purpose of the above field-redefinition `trick' is to simplify the calculation, but 
it also explains why Metz and Drechsel~\cite{Metz:1996fn}, calculating
polarizabilities in the linear sigma model with heavy $\si$-meson,
obtain to one loop exactly the same result as B$\chi$PT at $\cO(p^3)$~\cite{Bernard:1991rq}. 
Indeed, we can see that the two calculations
are related through a field redefinition and therefore bound to give the same results for
physical quantities. 

A word on renormalization. The one-particle-reducible graphs in 
Figs.~\fref{loops} and \fref{loopsD} contribute to the nucleon 
mass, field, charge, and magnetic moment renormalization. We have adopted the
on-mass-shell renormalization scheme, and used then the following values
of the parameters:  $e^2/4\pi = 1/137$, $g_A=1.267$, $f=f_\pi = 0.0924$ GeV,
$m_\pi = 0.139$ GeV,
 $M_N = 0.9383$ GeV, $\kappa_N = 1.79$ for the proton.
 
 The $\De$ contribution is included here within the framework
developed in Ref.~\cite{Pascalutsa:2003zk}. A study of the $\De$-resonance contributions
is necessary before embarking on to the $p^4$ calculations, since some of the $\Delta$-contributions are
of order $p^4/\vDe$, where $\vDe=M_\De-M_N\approx 0.3$ GeV. The diagrams that contribute at that
order are shown in \Figref{loopsD}. The first graph, the $\Delta$ Born contribution has been
 calculated in the same way as in Ref.~\cite{Pascalutsa:2003zk},
 except that now we have used the values of the $\ga N\to \De$ transition parameters ($g_M=2.95$ and $g_E=-1.0$) from
the pion-photoproduction analyses of Refs.~\cite{Pascalutsa:2005ts,Pascalutsa:2006up}, and 
also included the corresponding crossed graph. For the $\pi N\De$ coupling we have used
$h_A=2.85$, the value inferred from the $\De \to \pi N$ decay width of 115 MeV.

\begin{figure}[tb]
\centerline{\epsfclipon   \epsfxsize=8.5cm%
  \epsffile{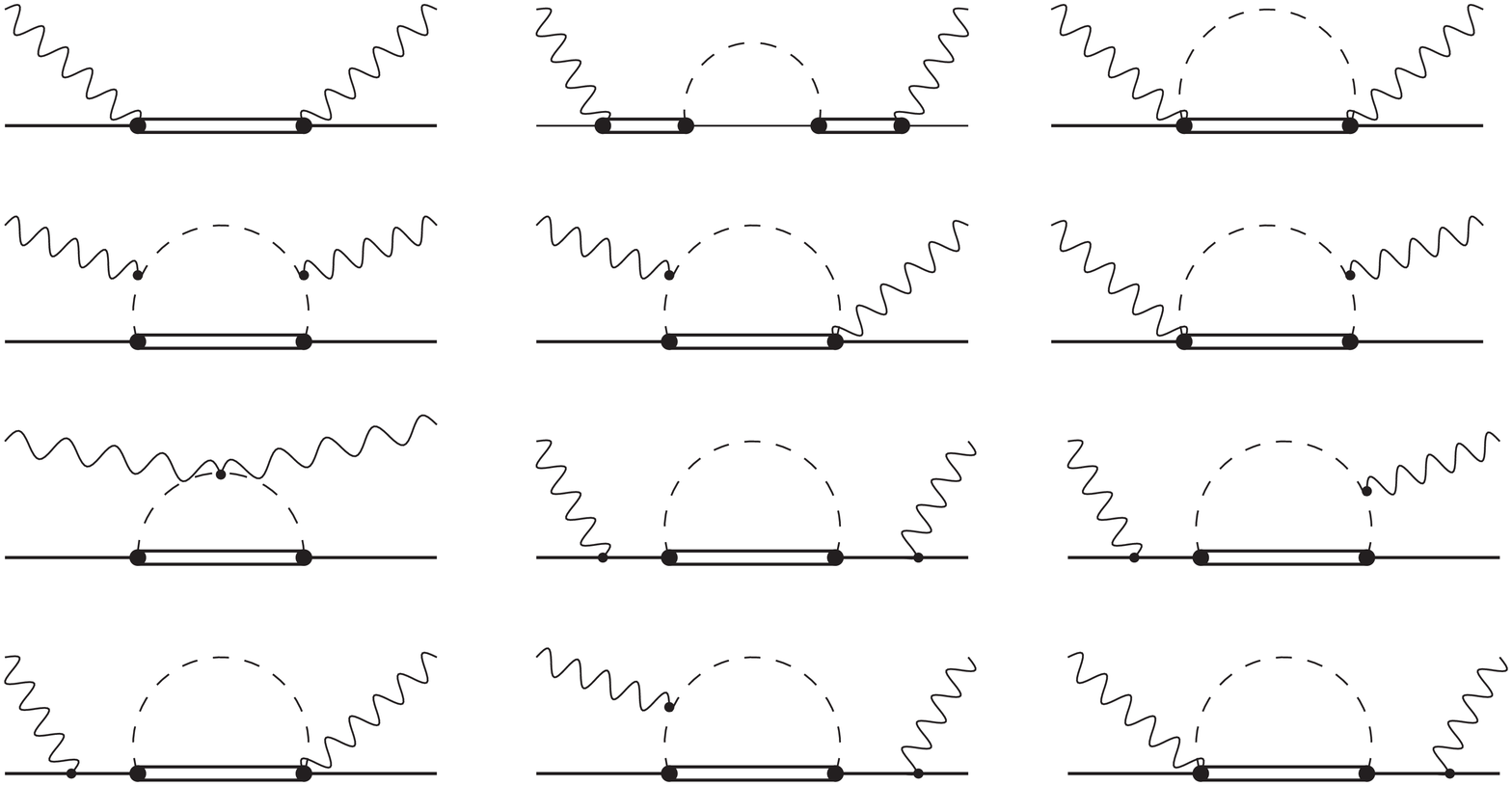} 
}
\caption{%(Color online)  
The graphs with $\Delta(1232)$ that contribute at order $p^4/\vDe$. 
Graphs obtained from these by
crossing are not shown, but are calculated as well.
}
\figlab{loopsD}
\end{figure}

The loop graphs in Figs.~\fref{loops} and \fref{loopsD} were eventually computed by us with the help of
algebraic tool FORM \cite{FORM} and the 
LoopTools library~\cite{Hahn:2006qw}. More details will be given in a subsequent publication.
In the rest of this letter we deal with
the results of these calculations.

\medskip

{\it 3. Consistency with sum rules}

The dispersion relations enjoy a special role in nucleon Compton scattering,
see Ref.~\cite{Drechsel:2002ar} for a nice review. To begin with,
practically all empirical values of nucleon polarizabilities 
are extracted from data with
the use of a model based on dispersion relations~\cite{Lvov:1980wp,L'vov:1996xd}.
Furthermore, in the forward kinematics, the Compton amplitude can be related to 
an integral over energy of the photoabsorption cross-section, which in combination
with the low-energy expansion yields a number of model-independent sum rules.
A famous example is the Baldin sum rule:
\beq
\alpha+\beta=\frac{1}{2\pi^2}\int\limits^\infty_0
 d\omega\frac{\sigma_T(\omega)}{\omega^{2}-i0}\eqlab{BSR},
\eeq
where the sum of polarizabilities is related
to an integral of the total photoabsorption cross-section $\sigma_T$.

In general, the forward Compton-scattering amplitude can be 
decomposed into two scalar functions of single variable in the following way:
\beq
T_{fi}(\omega)=\vec\epsilon^{\,\prime*}\cdot\vec\epsilon \,f(\omega)
+i\vec\sigma\cdot(\vec\epsilon^{\,\prime*}\times\vec\epsilon\,)\,\w\,g(\omega),
\eeq
where $\vec\epsilon^{\,\prime},\ \vec\epsilon$ are the polarization vectors of the initial and final photons, respectively,
and $\vec\sigma$ are the Pauli spin matrices. The functions $f$ and $g$ are
even functions of the photon energy $\w$. 
Using analyticity and  the optical theorem (unitarity)
one can write down the following 
sum rules:
\begin{subequations}
\bea
f(\omega) &=& f(0)+\frac{\omega^2}{2\pi^2}\int\limits^\infty_0
 d\omega^\prime\frac{\sigma_T(\omega^\prime)}{\omega^{\prime\,2}-\omega^2-i0}\eqlab{fdis}\,,\\
g(\omega) &=& \frac{1}{4\pi^2}\int\limits^\infty_0
 d\omega^\prime\,\omega^\prime\, 
\frac{\sigma_{1/2}(\omega^\prime)-\sigma_{3/2}(\omega^\prime)}{\omega^{\prime\,2}-\omega^2-i0}\eqlab{gdis}\,,
\eea
\eqlab{sumrules}
\end{subequations}
where $\sigma_\lambda$ is
the doubly-polarized photoabsorption cross-section, with $\la$ being the helicity
of the initial photon-nucleon state.

These sum rules should also hold for the individual contributions of the
loop graphs in \Figref{loops}. In this case 
the photoabsorption process is given by the leading-order  single-pion
photoproduction, for which analytic expressions 
exist~\cite{Holstein:2005db}. We were able to verify that 
the (renormalized) $p^3$ loop contributions in \Figref{loops} 
fulfill the sum rules \Eqref{sumrules} exactly.

It is interesting to note that the leading-order pion photoproduction amplitude,
which enters on the right-hand side of \Eqref{sumrules}, is independent of
whether one uses pseudovector or pseudoscalar $\pi NN$ coupling. 
It essentially means that chiral symmetry is not relevant at this order. 
The latter statement can, by means of the sum rule, be extended 
to the forward Compton amplitude
at $\cO(p^3)$. On the other hand, the graphs $(h)$ and $(i)$ in \Figref{loops},
being the only ones beyond the pseudoscalar theory, take the sole role
of chiral symmetry. In the forward kinematics these graphs indeed vanish,
but play an important role in the backward angles. Without them the values of $\al$ and $\be$ would be entirely different.
The value of $\al+\be$ would of course be the same, but $\al -\be$
would (approximately) flip the sign. Furthermore, in the chiral limit, the value of
$\al -\be$ would diverge as $1/m_\pi^{2}$ (instead of $1/m_\pi$ as it should). 
In this way we arrive at the conclusion that chiral symmetry 
plays a more prominent
role in the backward Compton scattering.

The resulting values for the polarizabilities, however, are in worse agreement
with experiment than the HB$\chi$PT $p^3$ result (see Table~\tabref{albe}).  
This, in fact, is a big success of the covariant calculation --- it opens a room for the
$\De$(1232)-resonance contributions. The $\De$-excitation plays 
an important role in nucleon polarizabilities, as can already be seen
from the Baldin sum rule and the fact that the 
photoabsorption cross-section is dominated, at lower energies,
by the $\De$ resonance. In B$\chi$PT we will see the $\De$ playing the same prominent
role on both sides of the sum rule. 
In contrast, the HB$\chi$PT $p^3$  value for $\alpha+\beta$ already saturates the sum rule, leaving
no room for other contributions.

\begin{figure}[bth]
\centerline{ \epsfclipon  
\epsfxsize=6.5cm%
  \epsffile{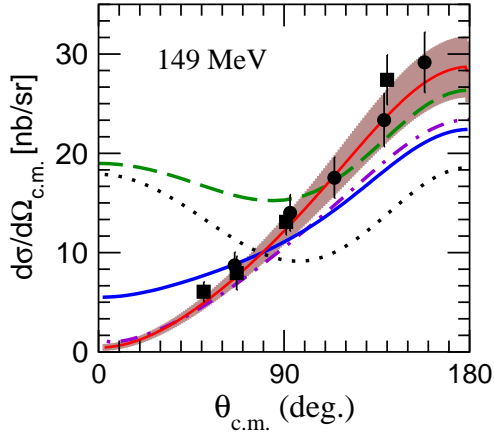} 
}
\caption{(Color online)  Angular dependence of 
the $\ga p\to \ga p$ differential cross-section in
the center-of-mass system for fixed photon-beam energy, 
$E_\gamma^{(lab)}=149$ MeV. Data points are from SAL~\cite{Hal93} ---
filled squares, and MAMI~\cite{MAMI01} --- filled circles. The curves are:
Klein-Nishina --- dotted, Born graphs and WZW-anomaly --- green dashed,
adding the HB$\chi$PT --- violet dash-dotted, adding the B$\chi$PT
--- blue solid. The result of adding the $\Delta$-excitation
contribution to the  B$\chi$PT $p^3$ is shown by the red solid line with a band.}
\figlab{fixE}
\end{figure}

\medskip

\begin{figure}[t]
%\vskip1mm
\centerline{\epsfclipon  \epsfxsize=6cm%
  \epsffile{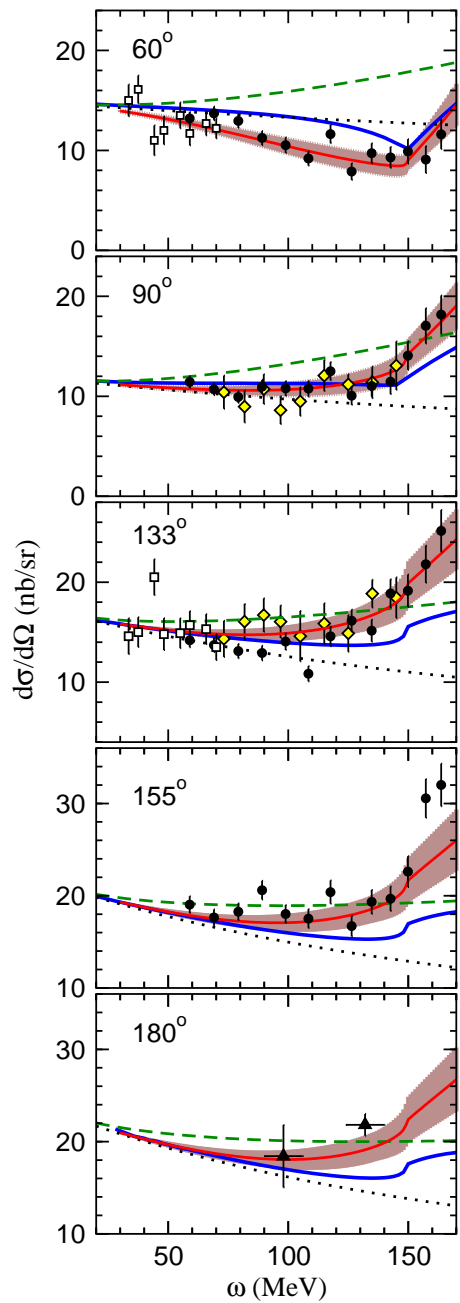} 
}
\caption{(Color online) Energy-dependence of 
the $\ga p\to \ga p$ differential cross-section in
the laboratory frame for fixed values of the scattering angle. 
Data points are from: Illinois~\cite{Federspiel:1991yd} --- open squares,
MAMI~\cite{Zieger:1992jq} --- filled triangles, SAL~\cite{MacG95} ---
open diamonds, and MAMI~\cite{MAMI01} --- filled circles. The legend
for the curves is the same as in \Figref{fixE}.} 
\figlab{fixA}
\end{figure}

{\it 4. Results for observables.}

First let us see how the differences
between the HB$\chi$PT and B$\chi$PT show up in  observables.
In \Figref{fixE}, we consider the unpolarized differential cross-section
of the $\gamma p\to\gamma p$ process 
as a function of the scattering angle in center-of-mass system, with
the incident photon energy fixed at just below the pion-production 
threshold. The dotted line represents the Klein-Nishina cross-section
(scattering off a classical pointlike `proton'). The dashed line adds on
the anomalous magnetic moment and the $\pi^0$ anomaly contributions. 
Then the effect of the $p^3$ HB$\chi$PT result is indicated by the
dash-dotted line, in contrast to the effect B$\chi$PT result 
which is given by the blue 
solid line.

Clearly, the major differences between the two $p^3$ calculations
arise at forward angles. This is because at low energies the 
$p^3$ contribution to the cross-section at forward (and backward) angles
is determined by the $p^3$ contribution to $\al+\be$ (and $\al-\be$).
We have seen already from Table~\tabref{albe}
that the sum of polarizabilities differs between the two calculations
much more than their difference, 
and this fact reflects itself in the cross-section.
 
The red solid line with an error band in \Figref{fixE}  shows the
result of adding the $\De$(1232)-resonance contribution to the covariant $p^3$
result. The fact that the $\De$ contribution in B$\chi$PT 
is compatible with both photoproduction
and Compton scattering data is further demonstrated 
in \Figref{fixA}, where the Compton scattering cross-section
is plotted as a function of photon energies at fixed angles (in the lab
system).  The legend for the curves is the same
as in the previous figure. The HB$\chi$PT result is omitted here, but
can be found in Ref.~\cite{Beane:2004ra}. 
The results for the nucleon proton polarizabilities, complete up to ${\cal O}(p^4/\vDe)$,
are displayed in the corresponding column of Table~\tabref{albe}.

We emphasize that,
while the fit in~\cite{Pascalutsa:2003zk}, where the HB$\chi$PT
$p^3$ result was used, demanded a re-adjustment of the $\ga N\De$ parameters
to unrealistic values, the present B$\chi$PT result 
allows for the correct values of the $\ga N\De$ parameters. 
Therefore up to the order $p^4/\vDe$ there no fitting parameters. This
is a prediction of B$\chi$PT.

\medskip

{\it 5. Conclusion}

We have studied of the nucleon Compton scattering in the
framework of B$\chi$PT at orders $p^3$ and $p^4/\vDe$. 
These are `predictive' calculations in the sense that there no new 
low-energy constants, they come starting at order $p^4$. 
We find that the covariant $p^3$ result fulfills the forward-Compton-scattering sum rules, while
the HB$\chi$PT result does not. Chiral symmetry has no effect on
the forward Compton scattering but plays an important 
role at the backward scattering.
Examining the Compton-scattering cross sections we find that the difference between the HB$\chi$PT and B$\chi$PT
results can indeed be unnaturally large, most notably in the forward kinematics. We argue that higher-order effects of the  $\Delta$(1232)-resonance
excitation can more naturally be accommodated in the manifestly covariant calculation.
This is due to partial cancellation of the relativistic and $\De$-excitation effects which
is explicit in the covariant calculation.
In contrast to the HB$\chi$PT approach, in B$\chi$PT the effect of $\Delta$(1232) appears to be
compatible both with the Compton scattering and pion photoproduction data. 

\medskip

{\it Acknowledgments}. 
We thank J.~Gegelia and M.~Vanderhaeghen for reading the manuscript
and useful remarks. 
This work is partially supported  by the European Community Research Infrastructure Activity under the FP6 
"Structuring the European Research Area" programme (HadronPhysics, contract RII3-CT-2004-506078).

\end{document}